\documentclass{iopart}

\usepackage{latexsym,graphicx} 
\begin{document}

\title[The effect of structure and geometry on vdW forces]{Numerical
study of the effect of structure and geometry on van der
Waals forces}

\author{C E Rom\'{a}n-Vel\'{a}zquez and Bo E Sernelius}

\address{Dept. of Physics, Chemistry, and Biology, Link{\"o}ping
University, SE-581 83 Link{\"o}ping, Sweden}
\ead{roman@ifm.liu.se}
\begin{abstract}
We use multipolar expansions to find the force on a gold coated sphere
above a gold substrate; we study both an empty gold shell and a gold 
coated polystyrene sphere.  We find four characteristic separation 
ranges.  
In the first region, which for the empty gold shell occurs for distances, 
$d$, smaller than the thickness of the coating, the result agrees with 
that on a solid gold sphere and varies as $d^{-2}$; for larger separations 
there is a region where the force behaves as if the coating is strictly 
two dimensional and varies as $d^{-5/2}$; in the third region the 
dependence is more unspecific; in the forth region when $d$ is larger than 
the radius, the force varies as $d^{-4}$.  For homogeneous objects of more 
general shapes we introduce a numerical method based on the solution of an 
integral equation for the electric field over a system of objects with 
arbitrary shapes. We study the effect of shape and orientation on the van 
der Waals interaction between an object and a substrate and between two 
objects.
\end{abstract}


\section{Introduction}
In his seminar paper \cite{casimir} Casimir calculated the energy between
two parallel perfect conductor half spaces as the change of zero-point
energy of the classical electromagnetic field,
\begin{equation}
\mathcal{U}\left(z\right)=\frac{\hbar
}{2}\sum_{i}\left[\omega_{i}\left(z\right)-\omega_{i}\left(z
\rightarrow\infty\right)\right],
\label{eqcasi} 
\end{equation}
where $\omega_{i}\left(z\right)$ are the frequencies of the electromagnetic
field modes of the plates separated by a distance $z$.  This procedure has
been employed by van Kampen \cite{kampen} to study vdW (van der Waals)
forces of dielectric plates in the non-retarded case.  Later Gerlach
\cite{gerla} interpreted the retarded Casimir forces in terms of
interacting surface plasmons of the plates.  Most recently the problem of
Casimir energy between two dielectric plates has been revisited and it has
been shown that it can be obtained as the sum of contributions from the
interacting surface plasmons and from the propagating modes of a cavity
formed by the parallel plates \cite{SerWiley, SerPRB, lambre}.  Apart from
the original results for parallel plates and slabs, a limited number of
results exists for other geometries and combination of materials.  The most
well known result is the so called proximity force approximation (PFA)
\cite{Derj}.  In that model the interaction between objects of various
shapes is calculated in terms of the energy between planar interfaces.  In
the case of a spherical object of radius $R$, above a planar substrate, the
force $F$ becomes
\begin{equation}
F\left(z\right)=-\frac{dV(z)}{dz}=2\pi
RE_{p}(z)\left[1-\frac{1}{RE_{p}(z)}\int_{z}^{z+R}dhE_{p}\left(h\right)
\right],
\label{pfa}
\end{equation}
where $z$ is the closest distance and $E_{p}(h)$ is the interaction energy 
per unit area between two planar interfaces a distance $h$ apart. The 
first factor of the result is what one usually means with the PFA. The 
second factor is a correction which we neglect in this work.
In the case of vdW forces other analytical results exist for highly 
symmetric geometries. One of them is the following general expression for 
the interaction energy between two cylinders, spheres or half spaces at 
small separations
\begin{equation}
E_\mathrm{vdW}=-A_{12}(\omega)\left[\frac{2\pi 
R_{1}R_{2}}{R_{1}+R_{2}}\right]^{1-n/2}\Gamma(1+n/2)L^{n}d^{-(1+n/2)},
\label{analitic}
\end{equation}
where $n=0$ for spheres, $n=1$ for cylinders, $n=2$ for half spaces and 
$A_{12}$ is the so called Hamaker constant given by
\begin{equation}
A_{12}(\omega)=\frac{\hbar}{32\pi^{2}}\sum_{l=1}^{\infty}\frac{1}{l^{3}}
\int_{-\infty}^{\infty}d\omega\left[\frac{\varepsilon_{1}(\omega)-
\varepsilon_{a}
(\omega)}{\varepsilon_{1}(\omega)+\varepsilon_{a}(\omega)}\frac
{\varepsilon_{2}
(\omega)-\varepsilon_{a}(\omega)}{\varepsilon_{2}(\omega)+\varepsilon_{a}
(\omega)}\right]^{l}.
\end{equation}
It has been shown that for larger separations, contrary to this formula,
the energy is not symmetric with the interchange of materials between the
objects \cite {nog2}.  The results of (\ref{analitic}) in fact coincide
with those obtained with the usual PFA approach for those geometries, in 
the nonretarded case which we are concerned with here.

Another set of results use additivity of pair interactions between the
atoms of two objets to obtain the behavior of the force at close and at
long distances.  In spite of the well known fact that the Casimir force is
not an additive interaction, there is a general consensus that these
results give the correct distance dependence of the forces.  One example is
that for two cylinders; for parallel cylinders the vdW energy goes like
$\sim d^{-1.5}$ when $d\rightarrow0$, but when they are crossed, a
dependence $\sim d^{-2}$ is expected, which is the same as between two
spheres.

A spectral representation formalism has been employed \cite{nog1} to obtain
the resonance frequencies of a sphere above a substrate from which one
obtains the vdW energy.  In the spectral representation formalism, the
optical response of a two face system is expressed as the sum of resonant
terms from which it is possible to explicitly obtain its resonance
frequencies \cite{repre}.  Those resonances correspond to the surface
plasmons of the system.  Furthermore, in that formalism the effects of
geometry and material are clearly separated; this makes it possible to
deduce the importance of the geometry in the resonances of the system, and
consequently in its vdW energy.

In this work we use a multipolar method similar to that in \cite{nog1} to
calculate the vdW interaction between a substrate at arbitrary separation
from a spherical object with an internal structure in the form of a sphere
with a coat of a different material.  We must mention that there are no
results for vdW interactions between finite objects with a structure formed
by parts of different materials.  We also introduce a formalism for the
determination of the vdW energy of a system of objects of arbitrary shape,
separation and orientation.  We present results for the vdW forces between
cubes, and finite cylinders above substrates, and develop calculations of
lateral and rotational forces between objects.
%
%
\section{Formalism}

\subsection{Multipolar method}
\begin{figure}
\center
\includegraphics[width=6cm]{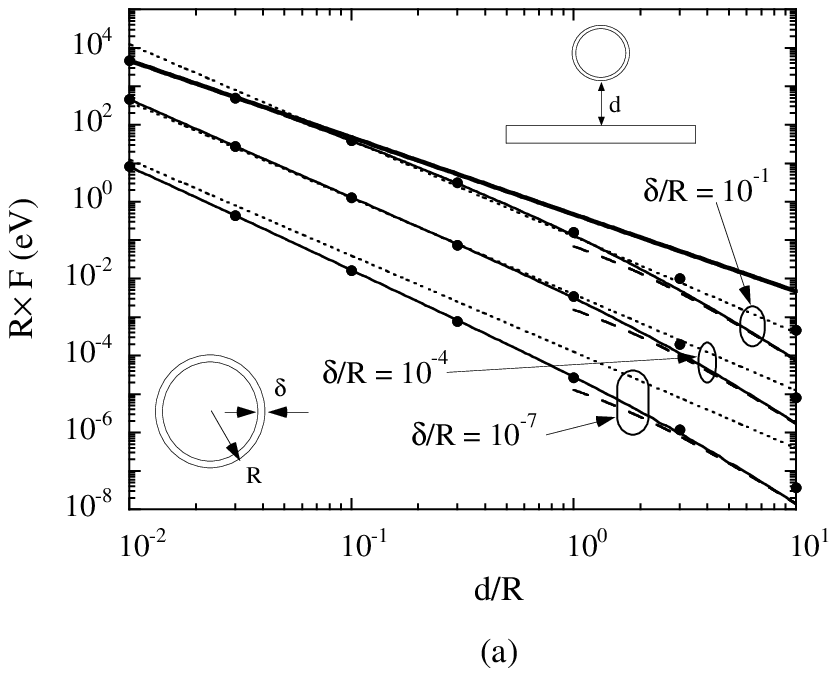}
\includegraphics[width=6cm]{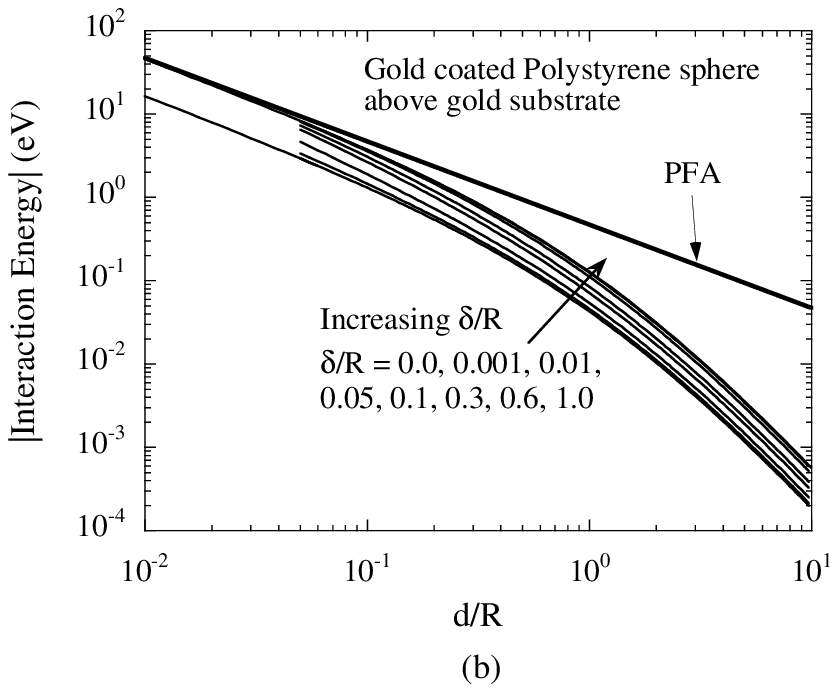}
\caption{(a) The radius times the force on a gold shell of radius $R$ and
thickness ${\delta}$ at distance $d$ above a gold substrate as function of
$d/R$.  See the text for details.  
(b) The interaction energy for a gold
coated polystyrene sphere of radius $R$ and coat thickness ${\delta}$ at
distance $d$ above a gold substrate as function of $d/R$.  The coat
thickness takes on the values 0.0, 0.001, 0.01, 0.05, 0.1, 0.3, 0.6 and 1,
in units of $R$, counted from below.  The thick solid curve is the PFA
result based on two gold half spaces.}
\label{figu1}
\end{figure}

First we consider the system shown in the inset of figure \ref{figu1}(a). 
A system of four materials bounded by two concentric spheres of radii
$R-\delta$ and $R$ with the center of the spheres at a distance $R+z$ from
the surface of a substrate.  The material of the interior of the sphere
with radius $R-\delta$ has a dielectric function $\varepsilon_{i}(\omega)$;
the coat or shell bounded by $R-\delta$ and $R$ contains a material with
dielectric function $\varepsilon_{c}(\omega)$; the substrate has a
dielectric function $\varepsilon_{s}(\omega)$. The full sphere is immersed
in an ambient with dielectric function $\varepsilon_{a}(\omega)$.  It has
been shown \cite{nog1,roman} that the multipolar moment $Q_{lm}$ of the
surface charge induced by the fluctuations of the electromagnetic field
with multipolar component $V_{lm}^{vac}$ of the stochastic vacuum field,
satisfies the system of linear equations
\begin{equation}
\sum_{l'}\left[\frac{1}{\alpha_{l}}\delta_{l,l'}+f_cA_{l,l'}^{m}(z)\right]
Q_{l'm}=-V_{lm}^{vac},\quad 
\left|m\right|\leq l,l' ,\label{eq1} 
\end{equation}
where $\alpha_l$ is the polarizability of the sphere (independent of $m$),
$f_c=\left[\epsilon_s(\omega)-\epsilon_a(\omega)\right]/\left[\epsilon_s(\omega)+\epsilon_a(\omega)\right]$,
and $A_{l,l'}^{m}$ are the terms, of the matrix that relate the multipolar
functions between the different systems of spherical coordinates of the
sphere and its image, that vanish when $z\rightarrow0$
\cite{roman,nog1}. For a homogeneous sphere $\alpha_l$ is given by
\begin{equation}
\alpha_{l}^m=-\frac{c_{l}}{u(\omega)-n_{l}},\quad c_{l}=la^{l+1/2}/(2l+1),
\quad n_{l+1}=l/(2l+1),
\label{eqpolari} 
\end{equation}
where $a$ is the radius of the sphere and
$u(\omega)=\left[1-\varepsilon_{sp}\left(\omega\right)/\varepsilon_{a}(\omega)\right]^{-1}$
is the so called spectral variable.  The function
$\varepsilon_{sp}(\omega)$ is the dielectric function of the sphere
material.  From (\ref{eq1}) and (\ref{eqpolari}) one can obtain a spectral
representation for $Q_{lm}$,
\begin{equation}
Q_{lm}=\sum_{l'}\left[\sum_{s}\frac{C_{s;l,l'}^{m}(z)}{u(\omega)-n_{s}^{m}
(z)}\right]V_{l',m}^{vac}.
\label{eqspectra} 
\end{equation}
The strengths $C_{s;l,l'}^m=\sqrt{c_lc_{l'}}U_{s,l}^{m}U_{s,l'}^{m}$ are 
the so called spectral functions, where $U_{s,l}^m$ is the orthogonal 
matrix that satisfies
\begin{equation}
\sum_{l,l'}U_{s,l}^{m}H_{l,l'}^{m}(z)U_{s',l'}^{m}=n_{s}^{m}(z)\delta_
{s,s'}. 
\label{eq5} 
\end{equation}
Here,
$H_{l,l'}^{m}(z)=n_{l}^{0}\delta_{ll'}+f_{c}\sqrt{c_{l}}\sqrt{c_{l'}}A_
{l,l'}^{m}(z)$ is a real symmetric matrix.  The resonance frequencies are
given by $u(\omega_{s}^{m})=n_{s}^{m}$, and the vdW energy is obtained from
(\ref{eq1}).  Note that the matrix $H_{l,l'}^{m}$ (and consequently $n_s$)
depends only on the geometrical properties of the system and the dielectric
properties of the substrate.  In the case of a coated sphere the
polarizability can be obtained using the boundary conditions of the
electric potential as
\begin{equation}
\fl
\alpha_{l}=-n_{l}R^{2l+1}\frac{\left[n_{l}-u_{ic}(\omega)\right]-\left[n_{l}-u_{ac}(\omega)\right](1-\delta/R)^{2l+1}}{\left[n_{l}-u_{ic}(\omega)\right]\left[n_{l}-u_{ca}(\omega)\right]+n_{l}(1-n_{l})(1-\delta/R)^{2l+1}}.
\end{equation}
with
$u_{xy}(\omega)=1/\left[1-\varepsilon_{x}(\omega)/\varepsilon_{y}(\omega)\right]$. 
Now it is neither possible to find a spectral representation of the system
nor to obtain its resonances explicitly.  One must rely on the argument
principle to calculate the energy of the system, as follows.  The
determinant of (\ref{eq1}) becomes zero at the resonance frequencies of the
system,
\begin{equation}
G(\omega_s,z)=\det\left[\frac{1}{\alpha_{l}(\omega_s)}\delta_{l,l'}+f_{c}A_{l,l'}^{m}(z)\right]=0.
\end{equation}
It is possible to apply the argument theorem to the multipolar problem
\cite{nog2} and using the fact that when $z\rightarrow\infty$ the
resonances are those of the isolated coated sphere to find that
(\ref{eqcasi}) is given by
\begin{equation}
\mathcal{U}\left(z\right)=
-\frac{\hbar}{i4\pi}\int_{-i\infty}^{i\infty}d\omega\log\left[G(\omega,z)/\prod_{l}\frac{1}{\alpha_{l}(\omega)}\right]. 
\label{eqargu}
\end{equation}
This expression can also be used in the case of experimental dielectric
functions or complex valued model dielectric functions for the sphere and 
substrate.

%
%
\subsection{Integral equation method}
The multipolar method above gives very high precision results over a wide
range of distances.  However to deal with particles of other shapes the
multipolar method becomes much more complex and expensive numerically. On
the other hand for finite objects of arbitrary geometry it is not possible
to define one unique way of calculating the energy in the PFA approach. 
For spheres and cylinders one considers small planes that interact with the
opposite in the direction parallel to the line that joins the centers of
the objects.  However for more complex geometries it is not clear how to
define the interacting planes. R Fuchs has developed a spectral
representation formalism for the calculation of the optical response, in
the non-retarded limit, of homogeneous objects of arbitrary shapes
\cite{spect}. We use his basic formalism for the formulation of a model to
obtain the vdW interaction energy for systems of objects of arbitrary
geometry.

Fuchs has shown that the surface charge
density, $\sigma$, satisfies the integral equation defined over the 
surface $\Sigma$ 
of the object
\cite{spect},
\begin{equation}
\mathbf{E}^{vac}(\mathbf{r},\omega)\cdot\hat{\mathbf{n}}(\mathbf{r})=
2\pi\frac{\varepsilon\left(\omega\right)+1}{\varepsilon\left(\omega
\right)-1}
\sigma\left(\mathbf{r}\right)-\int_{\Sigma}\frac{\hat{\mathbf{n}}(\mathbf
{r})
\cdot(\mathbf{r}-\mathbf{r}')}{\left|\mathbf{r}-\mathbf{r}'\right|^{3}}
\sigma\left(\mathbf{r}\right),
\label{eqinte} 
\end{equation} 
where $\mathbf{E}^{vac}$ is the stochastic vacuum field that excites the 
particle and 
$\hat{\mathbf{n}}$ is the surface normal pointing outwards from the 
surface. The complete deduction of (\ref{eqinte}) is independent 
of the form and connectivity of the surface \cite{spect}. We use this 
integral 
equation when $\Sigma$ is the combination of the surfaces of two finite 
arbitrary objects. The simplest numerical solution of (\ref
{eqinte}) is obtained, as in \cite{spect}, by dividing the surface $\Sigma
$ in small fragments of area $\Delta s_i$ and surface normal 
$\hat{\mathbf{n}}_{i}$ and with homogeneous surface charge 
$\sigma_i$, centered at the points $\mathbf{r}_i$.  As a result, one 
obtains the linear system of equations, 
\begin{equation}
\fl
2\pi\frac{\varepsilon\left(\omega\right)+1}{\varepsilon\left(\omega
\right)-1}\sigma_{i}-\sum_{j}R_{i,j}\sigma_{i}=f_{i}, \quad
R_{i,j}=(1-\delta_{i,j})\frac{\hat{\mathbf{n}}_{i}\cdot(\mathbf{r}_{i}-
\mathbf{r}_{j})\Delta 
s_{i}}{\left|\mathbf{r}_{i}-\mathbf{r}_{j}\right|^{3}},
\label{fuchs}
\end{equation}
and $f_{i}=\mathbf{E}^{0}(\mathbf{r}_{i})\cdot\hat{\mathbf{n}}
(\mathbf{r}_{i})$.
A spectral representation can be obtained from solving (\ref{fuchs}),
\begin{equation}
\sigma_{i}=-\sum_{j,s}\frac{C_{s}^{i,j}}{u(\omega)-n_{s}}f_{j},
\end{equation}
where $n_s=(1-m_s/2\pi$)/2, $m_s$ are the proper values of the matrix 
$R_{i,j}$, and $C_s^{i,j}$ are obtained from the matrix that diagonalizes 
$R_{i,j}$ \cite{spect}. The resonance frequencies are given by $u(\omega_s)
=n_s$ and like in the case of the multipolar spectral representation of a 
sphere above a substrate, the vdW energy could be calculated 
with (\ref{eq1}).
%
%
\section{Results and Discussion}
\subsection{Free standing spherical shells and gold coated 
polystyrene spheres
\label{Coated}}
For the calculations in this part we first construct the matrix of the
linear system, (\ref{eq1}), for a given $z/R$.  For the dielectric function
of gold and polystyrene we use those in \cite{goldlamb} and
\cite{isra}, respectively.  We calculate numerically the determinant and
perform the integration of (\ref{eqargu}).  An increasing maximum number
$L_{max}$ ($l,l'< L_{max}$) of multipoles is considered until the
convergency of the energy is reached.

In figure \ref{figu1}(a) we show the results for a free standing gold
shell above a solid gold substrate.  As long as one may neglect retardation
effects the figure is universal, independent of the size of the sphere. 
The thick solid curve is the PFA result when $E_p(z)$, of (\ref
{pfa}), is the interaction energy between two half spaces; here the force
varies as $d^{ - 2} $.  We see from the figure that the full result from
using the multipolar method, thin solid curves, merge with the thick solid
curve for small enough separations.  This only happens inside the figure
for the thickest of the coatings.  For the other two examples it happens
for smaller separations.  For larger separations there is a region where
the result follows the PFA result when $E_p (z)$ is the interaction energy
between one half space and one strictly two dimensional film, dotted
curves; here the force varies as $d^{ - 5/2}$, i.e., has a fractional power
dependence.  The dotted curves are obtained as \cite{SerBjo,BoSer}
\begin{equation}
 R \times F = 2\pi R^2 E_p \left( d \right) \approx 0.1556\sqrt {{{n\hbar
 ^2 e^2 } \mathord{\left/ {\vphantom {{n\hbar ^2 e^2 } {m_e }}} \right. 
 \kern-\nulldelimiterspace} {m_e }}} {{\sqrt {{\delta \mathord{\left/
 {\vphantom {\delta R}} \right.  \kern-\nulldelimiterspace} R}} }
 \mathord{\left/ {\vphantom {{\sqrt {{\delta \mathord{\left/ {\vphantom
 {\delta R}} \right.  \kern-\nulldelimiterspace} R}} } {\left( {{d
 \mathord{\left/ {\vphantom {d R}} \right.  \kern-\nulldelimiterspace} R}}
 \right)^{{5 \mathord{\left/ {\vphantom {5 2}} \right. 
 \kern-\nulldelimiterspace} 2}} }}} \right.  \kern-\nulldelimiterspace}
 {\left( {{d \mathord{\left/ {\vphantom {d R}} \right. 
 \kern-\nulldelimiterspace} R}} \right)^{{5 \mathord{\left/ {\vphantom {5
 2}} \right.  \kern-\nulldelimiterspace} 2}} }}.
\end{equation}
The circles are the PFA result when $E_p (z)$ is the interaction energy
between a half space and a film of thickness ${\delta}$.  Note that the
last two calculations (dotted lines and circles) are extensions of the PFA
model; in the original model one used $E_p (z)$ of two half spaces. The
dashed curves are the result from only including dipolar interactions.

Measurements of the vdW and Casimir interactions often involve spheres
since in this case the difficult problem of alignment is obsolete.  In most
cases these spheres are not solid but coated; one assumes that the coated
spheres may represent solid spheres.  As we have seen, this assumption is
only valid for relatively small distances.  In figure \ref{figu1}(b) we
show the interaction energy for a gold coated polystyrene sphere above a
gold substrate.  This way of displaying the results also provides universal
curves, independent of the radius of the sphere.  The radius, $R$, is the
value for the full sphere including the coating.  We see that only the
spheres with very thick coatings behave as solid gold spheres.

\begin{figure}
\center
\includegraphics[width=8cm]{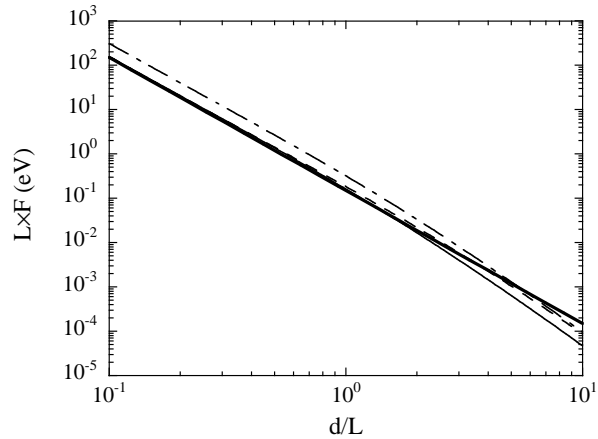}
\caption{vdW forces between gold prisms the distance $d$ above a gold
substrate.  The thin solid curve is the result for a gold cube of side $L$.
The dashed (dash-dotted) curve is the result for a standing up (lying down)
prism of length $2L$. The thick solid curve is the PFA result for the cube,
or standing up prism, based on two gold half spaces.}
\label{figu2}
\end{figure}
%
\subsection{Interaction between objects of arbitrary geometry}
\subsubsection{Objects above substrates:}
For the calculation of the vdW energy of these systems we define a
discretization of the surface of the object in small areas specified by
their positions, normal vectors and sizes.  With that information we
construct the matrix $R_{i,j}$ in (\ref{fuchs}) and obtain the eigenvalues
$m_s$ numerically and from them the resonance frequencies.  Finally, the
energy is calculated with (\ref{eq1}).  Using a maximum of 1000 discrete
elements per object we reached down to distances of 2\% of its maximum
diameter, within a numerical error of less than $2\%$, as estimated from
comparisons with calculations of \cite{nog2,nog1}.

In figure \ref{figu2} we show calculations for different prisms above
substrates, of the same material.  The thin solid line shows the result for
a cube of side $L$ at a distance $d$; the dashed line shows the result for
a prism of height $2L$ and square base with side length $L$; the thick
solid line shows the result of (\ref{analitic}).  We note that when the
cube and the square prism are very close to the substrate the force equals
the base area times the force per unit area between two half spaces.  At
long distance the force is dominated by dipolar interactions which are
proportional to the volume.  The dash-dotted line shows the result when the
prism is lying down, i.e., one of the rectangular sides is parallel to the
substrate.  So for small separations the force on the cube and standing up
prism are equal while it is twice the size on the lying down prism; for
large separations the forces on the standing up and lying down prisms are
equal but the force on the cube is only half the size.  The result for a
cylinder with height $L$ and a circular base with radius $R = L/\sqrt{\pi}$
follows closely the result for the cube, with a deviation smaller than the
numerical error $\sim 1\%$.  These results show that the shape of the side
that faces the substrate is not important.  One consequence of this result
is the additivity of the vdW energy for prisms: At all distances the vdW 
energy for prisms of same height and arbitrary shape is approximately 
proportional to the area of its base.

\begin{figure}
\center
\includegraphics[width=6cm]{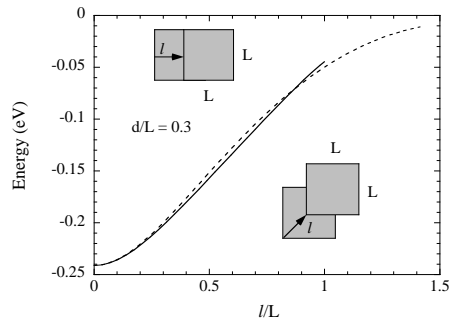}
\caption{The lateral energy for two gold square-based prisms.}
\label{figu3}
\end{figure}
%
\subsubsection{Interactions between two objects:}
The dependence of  the vdW interaction on the object shapes becomes more 
complex
in the case of the interaction between two objects.  The interaction tends
to modify the position and orientation of the particles to minimize the
energy.  It is possible to consider configurations in which two important
kinds of interaction arise: lateral forces and rotational forces.

In figure \ref{figu3} we show results for lateral forces between two prisms
of height $L/2$ with square bases of side $L$; they are placed above each
other at a fixed distance $d$; they are displaced in the horizontal plane a
distance $l$ from the complete parallel alignment.  The solid line shows
the results when the prisms are displaced in the direction parallel to one
of its side faces.  The dashed line shows the result when the prisms are
displaced in the diagonal direction relative its base (see the lower inset
of figure \ref{figu3}.)  We note that both curves reach a minimum in the
position of complete alignment.  From this one can expect an oscillatory
behaviour around this position.  Even thought one could expect the
additivity of the interaction in the limit when the particles are close to
each other, there is no such additivity with the conditions considered
here.  In the diagonal displacement the area of interaction increases
differently from that in the other type of displacement.  However the
energy shows an increase similar to that for the other displacement.  This
result suggests that the surface and boundary effects due to the
interaction between the surface plasmons of the different objects is more
important in the generation of lateral forces than in the case of an object
above a substrate.
\begin{figure}
\center
\includegraphics[width=6cm]{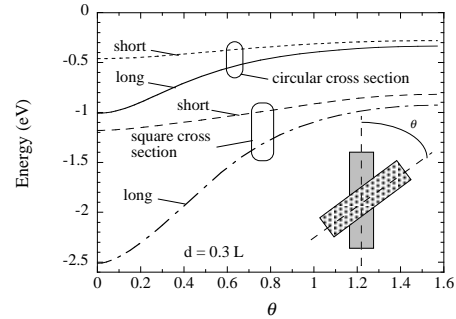}
\caption{The rotation energy for finite gold cylinders the distance $d=0.3L$ 
apart.}
\label{figu4}
\end{figure}

In figure \ref{figu4} we show results for the rotational energy between two
crossed cylinders as shown in the inset.  The cylinders are rotated
relative each other an angle $\theta$ around an axis passing through their
centers. The angle varies from $0$ to $\pi/2$.  The upper two curves are
for cylinders with circular cross section with diameter $L$.  The dotted
(solid) curve is for cylinders of length $L$ ($2L$).  The lower two curves
are for cylinders with square cross section of side length $L$.  The dashed
(dash-dotted) curve is for cylinders of length $L$ ($2L$).  In all cases
the closest distance, $d$, is $0.3 L$.

We see that in the case of perpendicular cylinders the interaction energy for
short and long cylinders are not the same, in disagreement with the
prediction from the PFA model.  This fact shows that the interaction
between the parts of the cylinder that do not face parts of the other one
has a strong contribution to the interaction energy.  In spite of the fact
that the interaction between the circular cylinders shows a change of
behaviour from that between two spheres, $\sim d^{-1}$, when they are fully
crossed to that between two parallel cylinders, $\sim d^{-3/2}$, when they
are aligned, we observe the strongest changes of energy in the case of
square cylinders; this is due to the fact that the plane surfaces have a
stronger interaction than the curved surfaces.  We can also see that there
is no perfect additivity in this configuration.  In the case of complete
alignment ($\theta = 0$) the cylinders of double length do not show double
energy.


\ack
This research was sponsored by EU within the EC-contract
No:012142-NANOCASE and we thank the organizing 
committee of the QFEXT07 for the partial support to attend the workshop.


\end{document}